\documentstyle[12pt]{article}
\begin{document}
\def\lsim{\raise0.3ex\hbox{$<$\kern-0.75em\raise-1.1ex\hbox{$\sim$}}}
\def\qsim{\raise0.3ex\hbox{$<$\kern-0.75em\raise-1.1ex\hbox{$q$}}}
\def\qsem{\raise0.3ex\hbox{$>$\kern-0.75em\raise-1.1ex\hbox{$q$}}}
\def\larr{\raise0.3ex\hbox{$\longrightarrow$\kern-1.5em\raise-1.1ex
\hbox{$\scriptstyle{q\rightarrow 1}$}}}
\def\rarr{\raise0.3ex\hbox{$\narray$\kern-1.5em\raise-1.1ex
\hbox{$\scriptstyle{N\rightarrow\infty}$}}}
\def\narray{\raise0.3ex\hbox{$\longrightarrow$\kern-0.8em\raise0.0ex
\hbox{${\scriptstyle /}$}}}
\def\qum{\raise0.3ex\hbox{$\sum'$\kern-1.5em\raise+2.4ex
\hbox{${\scriptstyle N/4}$\kern-1.5em\raise-3.9ex
\hbox{${\scriptstyle n_{k_1}\cdots n_{k_m}=0}$}}}}
\newcommand{\be}{\begin{eqnarray}}
\newcommand{\ee}{\end{eqnarray}}
\begin{titlepage}
\begin{flushright}
{\bf HU-SEFT R 1994-05}
\end{flushright}
\vskip 1 cm
\begin{center}
{\Large\bf New Phenomenon of Nonlinear Regge Trajectory and Quantum Dual
String Theory}
\vskip 1.0 cm
M. Chaichian\renewcommand{\thefootnote}{*}\footnote{Laboratory of High
Energy Physics , Department of Physics, P.O. Box 9
(Siltavuorenpenger 20C) FIN-00014, University of Helsinki, Finland}$^,$
\renewcommand{\thefootnote}{\dagger}\footnote{
Research Institute for High Energy Physics, P.O. Box 9 (Siltavuorenpenger
20C) FIN-00014, University of Helsinki, Finland} ,
J.F. Gomes\renewcommand{\thefootnote}{\dagger\dagger}\footnote{Instituto de
 F\'\i sica
Te\'orica-UNESP, Rua Pamplona 145, 01405-900 S\~ao Paulo, SP, Brazil} and
R. Gonzalez Felipe$^*$
\end{center}
\vskip 4.0 cm
\begin{abstract}

The relation between the spin and the mass of an infinite number of
particles in a $q$-deformed dual string theory is studied. For the deformation
parameter $q$ a root of unity, in addition to the relation of such values of
$q$ with the rational conformal field theory, the Fock  space of each
oscillator mode in the Fubini-Veneziano operator formulation becomes truncated.
Thus, based on general physical grounds, the resulting spin-(mass)$^2$
relation is expected to be below the usual linear trajectory. For such
specific values of $q$, we find that the linear Regge
trajectory turns into a square-root trajectory as the mass increases.
\end{abstract}
\end{titlepage}
As is well known string theory provides a promising approach to describe the
different forces and particles observed in nature. The development of this
theory is, more or less, directly related to the Veneziano's  discovery
\cite{vene} of a four-point crossing-symmetric scattering amplitude with
linear Regge trajectories. Afterwards Fubini, Gordon and Veneziano \cite{fubi}
provided an elegant operator formalism for the dual amplitudes, which
further led to the interpretation of the Veneziano model as a theory of
interacting strings in physical space-time.

In recent years
the mathematical structure of the quantum groups (see, e.g. \cite{fadd})
has been extensively explored
in connection with several important aspects of physical phenomena.
Among them, the classification of two dimensional conformally invariant field
theories plays an important role in understanding the structure of string
theory (see, e.g. \cite{green}). In particular, for rational conformal
field theories, described by a finite number of primary fields, the fundamental
properties of fusing and braiding of conformal blocks can be casted within the
product of representations of $q$-deformed algebras for the specific values
of the deformation parameter $q$ being a root of unity \cite{alva}.
For such values of $q$ quantum algebras
exhibit rich representation behaviours and are important in two-dimensional
conformal field theories and in statistical mechanics models \cite{vega}.

In Ref. \cite{chaic} the operator
formalism proposed in \cite{fubi} was followed to present
a $q$-deformed dual string model which possesses crossing symmetry in
exchanging
the $s$- and $t$-channels and factorization of the dual amplitudes in such
a way that they can be constructed as a field theory with Feynman-like
diagrams out of
products of vertices and propagators as the building blocks. However, as we
shall see below, the $q$-deformed dual amplitude proposed there leads to a
relation between the spin and the mass
spectrum of the particles which may not be of physical interest and in
principle could be rule out by experimental results.

In this letter we  propose
an alternative $q$-deformed dual string amplitude, which has not only
 the required properties of crossing symmetry and
factorization, the correct pole structure and a suitable asymptotic behaviour,
but also leads to a spin-(mass)$^2$ relation, which in our opinion possesses
features more interesting from the physical point of view. As in \cite{chaic}
we shall
introduce an infinite number of $q$-oscillators which build up a Fock
space with well-known properties. However, for the specific values of the
deformation parameter $q$ being a root of unity, only a finite number of
oscillators for each (harmonic) mode possesses non-vanishing norm.
This is a fundamental property which is responsible for a drastic change in
the high energy behaviour of
the amplitudes and, therefore, in the mass spectrum of the physical particles.
 Earlier attempts by introducing a logarithmic
behaviour for the Regge trajectories with  only a finite number
of oscillators \cite{coon} or by replacing the ordinary gamma functions by
their $q$-analogs \cite{roman} have been proposed. However, they cannot bring
to a field theoretical formulation of the dual string model.

Let us consider first the usual Veneziano \cite{vene} 4-point dual amplitude
given by
\be
A_4=\int^1_0 z^{-\alpha(t)-1} (1-z)^{-\alpha(s)-1} dz ,
\label{1}
\ee
where $\alpha(s) = \alpha^\prime s + \alpha_0$ is the linear Regge trajectory,
with $\alpha^\prime$ and $\alpha_0$ the Regge slope and intercept,respectively.
We will take our units so that $\alpha^\prime = 1$.
The amplitude (\ref{1}) can be factorized by introducing  an infinite set
of oscillators \cite{fubi} satisfying
$[a^\mu_m,a^{\nu^+}_n]=\delta_{mn}g_{\mu\nu}$ ; $\mu$ and $\nu$ are Lorentz
indices; $n,m = 1,2,\ldots,\infty$ correspond to the different oscillator
modes. We have then the identity
\be
(1-z)^{-A\bar{A}}=\prod\limits^\infty_{n=1}\langle 0\vert e^{\frac{Aa_n}
{\sqrt{n}}}\
z^{na_n^+a_n}\ e^{\frac{\bar{A}a_n^+}{\sqrt{n}}}\vert 0\rangle \ ,
\label{2}
\ee
where the contraction of the
Lorentz indices is understood in all the scalar products. Here the
 four-vectors $A_\mu$ and $\bar{A}_\mu$ correspond
 to incoming and outgoing momenta respectively. In what follows we shall denote
$a \equiv \alpha(s)+1 $ and $b \equiv \alpha(t)+1$ and omit the Lorentz
indices.

To perform the $q$-deformation within the operator formalism, it seems most
natural
to use the $q$-deformed oscillators \cite{macf} instead of the usual ones
 in the
factorization procedure, as this was done in \cite{chaic}, since
$q$-oscillators have many features in common with the usual harmonic
oscillators and from the quantum group point of view are the straightforward
generalization of the latters. Therefore, we will replace
the identity (\ref{2})  by the $q$-deformed expression
\be
F(a,z)=\prod\limits^\infty_{n=1}\;_q\langle 0\vert e_q^{\frac{Aa_n}{\sqrt{n}}}
z^{nN_n}\ e_q^{\frac{\bar{A}a^+_n}{\sqrt{n}}}\vert 0 \rangle_q =
\prod\limits^\infty_{n=1}\sum\limits^\infty_{\ell=0}
\biggl(\frac{a}{n}\biggr)^\ell\frac{z^{n\ell}}{[\ell]!}=
\prod\limits^\infty_{n=1}e^{\frac{a}{n}z^n}_q,
\label{3}
\ee
where $[\ell]=(q^\ell-q^{-\ell})/(q-q^{-1})$ , $[\ell]!=[1][2]\cdots[\ell]$
and $[0]!=1$;
$e^x_q=\sum\limits^\infty_{\ell=0}\frac{x^\ell}{[\ell]!}$ is the $q$-
exponential function.  The
$q$-oscillators entering in (\ref{3}) satisfy \cite{macf}
\be
a_ma^+_m-q a^+_ma_m=q^{-N_m}\ ,\ [N_m,a^+_m]=a^+_m\ ,\ [N_m,a_m]=-a_m,
\label{4}
\ee
with all the other {\it commutation relations} corresponding to different
indices of oscillators, vanishing. The operator $N_m$ is the
number operator corresponding to the mode $m$.

The main difference between the expression (\ref{3}) and the corresponding one
proposed in \cite{chaic} (cf. Eq. (8) in \cite{chaic}) is that we have
assumed the total energy operator of the system to be
$H=\sum\limits^\infty_{n=1}nN_n$ instead of
$H=\sum\limits^\infty_{n=1}na^+_na_n$
and therefore, we have
replaced $a^+_n a_n$ by the number operator $N_n$ in the exponent of $z$. As a
consequence of this assumption, the amplitudes defined in the present model
will exhibit poles in $s$ when $\alpha (s)$ is a positive
 integer, or poles in $t$ when $\alpha (t)$ is a positive integer as in the
undeformed case. For the amplitudes defined in \cite{chaic}, as we shall
see below, the poles are in general located at non-integer values.

In order to preserve duality, expressed by the symmetry $z\rightarrow 1-z$ in
(\ref{1}), we define then the $q$-deformed 4-point amplitude as
\be
A^q_4=\int^1_0F(a,z)F(b,1-z)dz=\prod\limits^\infty_{n=1}\;_q\langle
0\vert e_q^{\frac{Aa_n}{\sqrt{n}}}\int^1_0 z^{nN_n}F(b,1-z)dz\
e_q^{\frac{\bar{A}a_n^+}{\sqrt{n}}}\vert 0\rangle_q \ \ .
\label{5}
\ee
Inserting a complete set of orthonormal states
$1=\sum\limits_\lambda\vert\lambda\rangle_q \;_q\langle\lambda\vert$,
with $\vert\lambda \rangle_q =
\prod\limits_i\ \frac{(a^+_i)^{\lambda_i}}{\sqrt{[\lambda_i]!}}\ \vert 0
\rangle_q$ we find $A^q_4$ in its factorized form
\be
A^q_4=\sum\limits_\lambda V(A,\lambda)D(\lambda,b)V^+(\bar{A},\lambda) \ ,
\label{6}
\ee
where
\be
V(A,\lambda)=\;_q\langle 0\vert\prod\limits^\infty_{n=1}\
e_q^{\frac{Aa_n}{\sqrt{n}}}\vert\lambda\rangle_q
\label{7}
\ee
is the vertex operator (with only one leg off shell) and
\be
D(\lambda,b)=\;_q\langle\lambda\vert\int^1_0 dz\ z\;^{\sum\limits^\infty
_{n=1}nN_n}F(b,1-z)\vert\lambda\rangle_q
\label{8}
\ee
is the propagator.

The other vertex operator $\Gamma$ with two legs off shell ,
which is needed for higher $n$-point
functions ($n \geq 5$), is given now by
\be
\Gamma(\lambda,\lambda',p,\bar{p})=\;_q\langle\lambda\vert\prod\limits^\infty_
{n=1}
e_q^{\frac{\bar{p}a^+_n}{\sqrt{n}}}e_q^{\frac{pa_n}{\sqrt{n}}}\ \vert\lambda'
\rangle_q \ ,
\label{9}
\ee
where $p=\bar{p}$ is the momentum of the unexcited leg in the vertex.
 The 5-point $q$-deformed amplitude, e.g., now can be written as a Feynman-like
diagram in terms of products of vertices and propagators in the tree
approximation:
\be
A^q_5=\sum\limits_{\lambda,\lambda'}V(A,\lambda)D(\lambda,b_1)
\Gamma(\lambda,\lambda',p,\bar{p})D(\lambda',b_2)V^+(A,\lambda'),
\label{10}
\ee
where
$b_1,b_2$ are the Mandelstam variables of the corresponding channels.

Notice that the vertices (\ref{7}) and (\ref{9}) and the propagator (\ref{8})
differ from those proposed in \cite{chaic} since there, two infinite sets of
independent
unphysical oscillators ($b_i,b^+_i,i=1,2$), which were just auxiliary and not
necessary, were introduced. The corresponding $q$-deformed $n$-point
amplitudes,
however, are identical whether one uses these auxiliary oscillators or not.
Notice also that (\ref{7}), (\ref{8}) and
(\ref{9}) lead to the usual undeformed expressions for the vertices and the
propagators in the limit
$q\rightarrow 1$. The latter feature was not present in the vertices and
propagators defined in \cite{chaic}.

Let us first examine the singularities of $A^q_4$ as a function of $s$ and
$t$. Since the integral in (\ref{5}) is symmetric in $s$ and $t$, we need only
to analyse the singularities in one of the two variables, e.g. in the variable
$t$. We first observe that near the singular point $z=0$ the integrand
$F(b,1-z)$ in (\ref{5}) can be approximated as
\be
F(b,1-z)\sim z^{-b} ,
\label{11}
\ee
and therefore the integral (\ref{5}) is not defined when $\alpha (t) \geq 0$
(or $\alpha (s) \geq 0$) . Following the standard procedure \cite{mand} we can
define it by analytic continuation to show that the amplitude $A^q_4$ exhibits
poles in $t$ at any integer. The residue at the $n$-th  pole in the
$t$-channel is a
polynomial in $s$ of degree $J \leq n$, which when decomposed into spherical
harmonics describes the exchange of a set of particles of spins $\leq J$.

Let us consider  now the effects of the deformation in the high
energy behaviour of $A^q_4$ . In the undeformed case
the behaviour of the 4-point amplitude (1) for $a$ large
can be easily obtained (see,e.g.,\cite{chaich}) from the contour integral,
\be
A_4\sim\oint\limits_{z=0}\frac{dz}{2\pi i}(1+az+\cdots + \frac{a(a+1)\cdots
(a+n-1)}{n!}z^n+\cdots)z^{-b} ,
\label{12}
\ee
leading to
\be
A_4\sim\frac{a(a+1)\cdots(a+b-2)}{\Gamma(b)}\sim\frac{a^{b-1}}{\Gamma(b)} ,
\label{13}
\ee
where $\Gamma$ is the gamma function.

In order to discuss the high energy behaviour of the $q$-deformed 4-point
amplitude (\ref{5}), we will follow a procedure similar to the one used for the
undeformed case , namely, we write $A^q_4$ as
the contour integral
\be
A^q_4\sim\oint\limits_{z=0}\frac{dz}{2\pi i}F(a,z)z^{-b} \ \ .
\label{14}
\ee

Consider now a generic term in the product (\ref{3}) which can be written as
\be
C_{n_1,\ldots, n_m} a^{n_1+n_2+\cdots+n_m}z^{n_1+2n_2+\cdots +mn_m},
\label{15}
\ee
where $m=1,2,\ldots,\infty $ and $C_{n_1,\ldots,n_m}$
are coefficients which depend on $q$.
Since we are interested in the high energy behaviour of (\ref{14}), i.e. in
the case when $a$ is large, we should pick in $F(a,z)$ the highest power of
$a$ which gives nonvanishing contribution to the contour integral (\ref{14}).
We are therefore led to solve the algebraic equation
\be
n_1+2n_2+\cdots +mn_m = b - 1,
\label{16}
\ee
with the condition that the spin
\be
J=n_1+n_2+\cdots +n_m   ,
\label{17}
\ee
which is the power of $a$ in (\ref{15}), takes its maximum value.

In what follows we shall consider two cases:

$(i)$ Assume that $q$ is real.
Since in this case the $n_i$ in (\ref{16}) range from 0 to $\infty$ , we obtain
that the maximum value of $J$ in (\ref{17}) will be given by $J=b-1$ when
$n_1=b-1, n_i=0$ for $i \geq 2$. Thus we obtain from (\ref{15}) that for high
energies
\be
A^q_4 \sim a^{b-1}.
\label{18}
\ee

We notice that the high $s$-behaviour of the 4-point q-deformed amplitude is
proportional to $s^t$, which leads to a mass spectrum with a linearly-rising
Regge trajectory as in the undeformed case.

$(ii)$
Let us consider now the case when  the deformation parameter $q$ is a $K$-th
root of
unity, i.e. $q=\exp (2i \pi /K)$. The $q$-analogs $[\ell]$ thus become $[\ell]=
\sin (2\pi \ell/K)/\sin (2\pi/K)$. We should point out here the main
difference with
respect to the undeformed case: due to the truncation of
the Fock  space for each oscillator mode in the Fubini-Veneziano
operator formulation, each term in the product (\ref{3}) consists
of a finite series ending at $\ell=\tilde{K}$, where
\be
\tilde{K}=\left\{ \begin{array}{ll}
                   K-1 ,   & \mbox{for $K$ odd} \\
                   K/2-1 , & \mbox{for $K$ even .}
                   \end{array}
          \right.
\label{19}
\ee

In this case $0 \leq n_i \leq \tilde{K}$ and we have two possibilities:

If $b \leq \tilde{K}+1$, then it is always possible to find a solution of
(\ref{16})
such that $n_1=b-1 \leq \tilde{K} $ and $n_i=0, i \geq 2$, which gives the
maximum value of $J$ in (\ref{17}). Then as before the high energy behaviour
will be given by eq. (\ref{18}) and, thus, {\it the trajectories will be
linear}.

If $b > \tilde{K}+1$,  it is easy to see that for a generic $m$ the highest
power of
$a$ in (\ref{15}) is
$J=m\tilde{K} $ and is obtained when all the $n_i$ are equal to their maximum
value $\tilde{K} $. Then according to (\ref{16}) $J$ will satisfy the
second-order algebraic equation
\be
\frac{J(J+\tilde{K} )}{2\tilde{K} }=b-1 ,
\label{20}
\ee
which has the positive solution
\be
J=\frac{\tilde{K} }{2}\left\{ -1 + \sqrt{1+\frac{8(b-1)}{\tilde{K} }} \right
\}.
\label{21}
\ee
Thus finally we obtain that for large values of $a$ and for $b > \tilde{K}+1$
\be
A^q_4 \sim a^J,
\label{22}
\ee
where $J$ is given by  (\ref{21}).

We notice now from (\ref{21}) and (\ref{22}) that the high $s$-behaviour of
the $q$-deformed amplitudes is proportional
to $s^{\sqrt{t}}$ which leads to the mass spectrum with a {\it square-root
 Regge trajectory} $\alpha(t)=\sqrt{t}+const$.
It appears, as was expected by physical arguments, that the crucial change
in the high energy
behaviour of $A^q_4$ occurs due to the truncation
in the series in (\ref{3}) and as mentioned before, is a direct
consequence of $q$ being a root of unity. This change can be understood by
the finite character of the Fock  space as a consequence of the assumed
values of $q$.

It is worth mentioning that Eq. (\ref{21}) gives the position of the poles
in the leading Regge trajectory for the values of the spin $J=m\tilde{K} $.
In general, for an arbitrary value of the spin $J=m\tilde{K} +r$
with $r=0,1,\ldots,
\tilde{K}-1$, it is possible to show that the poles of the amplitude
are located at the
points given by the solution of the equation
\be
\frac{(J+r)(J+\tilde{K} -r)}{2\tilde{K} }=\frac{(2J-m\tilde{K} )
(m+1)}{2} = b-1 \ \ .
\label{23}
\ee

In Fig. 1 the leading Regge trajectories for both cases, when $q$ is real and
$q=\exp (2i \pi /K)$, are shown .
The dots denote the position of the poles and the
linearly-rising Regge trajectory  corresponds
to the case when the deformation parameter $q$ is real. We observe that
when $q$  is a $K$-th root of
unity, the trajectory is linear for $t \leq \tilde{K}$ and turns into  a
square-root trajectory for $t > \tilde{K}$, with $\tilde{K}$ given by Eq.(19).
 We also notice that for smaller
values of the parameter $K$ (i.e. when the truncation of the series
in (\ref{3}) occurs earlier), the effect of the deformation is enhanced. As
$K \rightarrow \infty$ , then $q \rightarrow 1$ and we recover the usual linear
trajectory. Since the amplitude is symmetric in $s$ and $t$, the resonances in
the $s$-channel also lie on the same Regge trajectory.

For comparison we shall consider the amplitude proposed in Ref. \cite{chaic}.
As we
already mentioned, the main difference consists on the choice of the
Hamiltonian
of the model, which in \cite{chaic} was taken as
$H=\sum\limits^\infty_{n=1}na^+_n a_n$ . In this case Eq. (\ref{3}) reads as
\be
F(a,z)=\prod\limits^\infty_{n=1}\sum\limits^\infty_{\ell=0}
\biggl(\frac{a}{n}\biggr)^\ell\frac{z^{n[\ell]}}{[\ell]!},
\label{24}
\ee
so that the deformation parameter $q$ enters explicitly in the exponent of $z$.
Therefore, in order to obtain the pole structure of the amplitude,
Eq.(\ref{16}) should be replaced by
\be
[n_1]+2[n_2]+\cdots +m[n_m] = b - 1,
\label{25}
\ee
 where as before the $n_i$ $ (i=1,2,\ldots,m) $ can take any integer value for
 $q$ real, and  $n_i \leq \tilde{K}$   when $q$
is a $K$-th root of unity.  It is possible to show that
in this case the poles of the $q$-deformed amplitude are defined by the
solutions
of Eq. (\ref{25}), which in general are given by real (and not only integer)
 values of $b$.  Nevertheless,the residue at each pole is a polynomial in
$a$ of degree $J$ given by Eq. (\ref{17}). Therefore, even when $q$ is real
 the Regge trajectories appear to be deformed in this case (Fig. 2).
  Another peculiarity of the amplitude defined in \cite{chaic}
appears in the case when
$q=\exp (2i \pi /K)$. Due to the relation $[K-\ell]=[\ell]$, there exists a
degeneracy in the solutions of Eq. (\ref{25}), which leads to a splitting
of the Regge trajectories, so that only spins of the form $J=m\tilde{K} $, with
$m$ being an integer number, lie on the leading trajectory.
In Fig. 3 we show the splitting of the trajectories for the particular
case of $K=3$, which is the same as for $K=6$. We see that only the poles
corresponding to even spins lie in
the leading Regge trajectory.  These features make the model proposed in
\cite{chaic} less appealing than the one presented here.

Thus quantum groups and the $q$-deformation provide with a new phenomenon of
a linear Regge trajectory to turn into a square-root trajectory for higher
masses in the case when the deformation parameter is a root of unity. This case
is of upmost physical interest and in particular, appears in rational conformal
field theories. An ultimate aim to combine the quantum group ideas with the
(super)string theory in order to obtain a $q$-deformed (super)string amplitude,
can be pursued along similar lines as presented in this letter.

\newpage
\noindent
{\bf Acknowledgements}
\vskip 0.5 cm
We are grateful to A. Demichev, L.A. Ferreira and A.H. Zimerman for
useful discussions. One of us (J.F.G.) thanks FAPESP and CAPES, Brazil, for
financial support and the Research Institute for High Energy Physics (SEFT),
University of Helsinki for the hospitality.

\vskip 1.5 cm
\noindent
{\large\bf Figure Captions}
\vskip 0.5 cm
\noindent
{\bf Fig. 1} \ \ The behaviour of the Regge trajectories for the $q$-deformed
dual string model
proposed in this letter. The dots denote the position of the poles and the
straight (dashed) line corresponds to the case when $q$ is
real, with linearly-rising Regge trajectory. When  $q$ is a $K$-th root of
unity, $q=\exp (2i \pi /K)$, the trajectory is linear for
$t \leq \tilde{K}$ and  turns into a square-root trajectory for $t >
\tilde{K}$, with $\tilde{K}=K-1$ for $K$ odd and $\tilde{K}=K/2-1$ for $K$
even (see Eq. (19)). The curves are shown for the particular values of
$K = 3, 4, 6, 8, 12$.
\vskip .5 cm
\noindent
{\bf Fig. 2} \ \ The Regge trajectories in the case of the model proposed in
\cite{chaic} for different values of
$q$ real.  The straight (dashed) line corresponds to the undeformed case. As
$q$ increases the deviation of the trajectories from the linear one
is enhanced.
\vskip 0.5 cm
\noindent
{\bf Fig. 3} \ \ Splitting of the Regge trajectories in the model of Ref.
\cite{chaic} for the case
when $q$ is a $K$-th root of unity. The trajectories are shown for the
particular
case of $K=3$, which is the same as for $K=6$. Only the poles
corresponding to even spins lie in the leading trajectory. The trajectories
behave as a square-root.

\end{document}